\documentclass[twocolumn,showpacs,showkeys,prd]{revtex4-1}

\usepackage{amsmath,amsfonts,amssymb,dsfont,mathrsfs,centernot}
\usepackage{graphicx}
\usepackage[colorlinks=True,linkcolor=blue,citecolor=blue]{hyperref}
\usepackage{comment}

\newcommand{\df}{{\bf{d}}}

\newcommand{\fy}{\centernot}

\newcommand{\ga}{\gamma}

\newcommand{\J}{\mathscr{J}}

\newcommand{\M}{\mathscr{M}}

\newcommand\VIN[2]{\hat{E}^{\hat{#1}}_{\hat{#2}}}

\newcommand\GAM[1]{{\gamma}^{\hat{#1}}}
\newcommand\gam[1]{\gamma^{{#1}}}

\newcommand\NAB[1]{\hat{\nabla}_{\hat{#1}}}

\newcommand\PA[1]{\partial_{\hat{#1}}}
\newcommand\pa[1]{\partial_{{#1}}}

\newcommand{\sgn}{\operatorname{sgn}}

\newcommand{\Ei}{\operatorname{Ei}}

\newcommand{\beq}{\begin{equation}}
\newcommand{\eeq}{\end{equation}}
\newcommand{\ber}{\begin{eqnarray}}
\newcommand{\eer}{\end{eqnarray}}

\renewcommand{\(}{\left(}
\renewcommand{\)}{\right)}

\hypersetup{pdftitle={On localisation of fermions on several domain wall models},pdfauthor={Oscar Castillo-Felisola},pdfkeywords={Branes, Domain Wall, Fermion Localisation, Moduli Space},pdflang={English}}

\begin{document}

\title{Localization of fermions in different domain wall models}
\author{Oscar Castillo-Felisola}
\email[E-mail address: ]{o.castillo.felisola@gmail.com}
\author{Iv\'an Schmidt}
\email[E-mail address: ]{ivan.schmidt@usm.cl}
\affiliation{Departamento de F\'\i sica y Centro Cient\'\i fico
Tecnol\'ogico de Valpara\'\i so, Universidad T\'ecnica Federico
Santa Mar\'\i a, Casilla 110-V, Valpara\'\i so, Chile.}

\begin{abstract}
  Localization of fermions is studied in different gravitational domain wall models. These are generalizations of the brane-world models considered by Randall and Sundrum, but which also allow gravitational localization. Therefore, they might be considered as possible realistic scenarios for phenomenology.
\end{abstract}


\maketitle

\section{Introduction}

In recent years, branes have been studied due to their uses in diverse topics of high energy physics. Those more relevant are: string dualities, which ended up with the breaking idea of the $AdS/CFT$ correspondence\cite{Maldacena1,Maldacena2}, and a new interpretation of compactification, which allows to postulate the $\M$-theory\cite{Horava:1995qa,Horava:1996ma} and the realization of brane-world scenarios\cite{RS1,RS2}.

These branes are extended objects arising in superstring theories, but their low energy analog might by realized through a gravitational domain wall.  Gravitational domain walls are solutions to the Einstein-Hilbert--Klein-Gordon equations where the self-interacting term of the scalar field has (at least) two degenerated minima and  the scalar field interpolates between consecutive  minima.

It has been shown that gravity does get localized on a $D3$-brane embedded in a higher dimensional  space-time,  in the sense that the effective theory of gravity on the domain wall is described by a Newtonian potential plus corrections, which coincide with the first one due to  general relativity.%%%%% A reference !!!
  However, in order to provide a  good phenomenological scenario, they should localize fermions and gauge bosons as well. Several generalizations to these models have been developed, such as considering more than two $D3$-branes\cite{Kogan:2000xc}, domain walls that are the thick versions of branes \cite{Gremm:1999pj}, domain walls generated by more than a single scalar field \cite{Almeida:2009jc}, and so on.

If thin 3-branes are considered, fermions cannot be localized \cite{Bajc:1999mh} and therefore, generalized brane-world scenarios have to be analyzed. Among the possibilities one might  generalize the brane to their thick analog. 

When a domain wall is considered, the simplest new interaction to be added is a Yukawa-like interaction between the scalar field which generates the domain wall and the fermions added to the formalism. However, in generic models only a single chirality is localized. In a previous work \cite{CastilloFelisola:2010xh}, the moduli space of parameter for a particular family of domain walls, known in the literature as double domain walls, was studied. It was found that there exist regions  where possibly both chiralities are localized, and that the most plausible scenario was highly dependent on the Yukawa-like coupling constant.

The aim of this work is to provide a similar analysis of the moduli space for diverse kinds of  domain walls, which is an essential ingredient in order to find out those that are acceptable candidates for phenomenology. In the next section a set of different domain wall solutions will be introduced. In section \ref{sec:loc} a brief review about localization of fermions on static branes  is considered. In the subsequent sections analysis and discussion are presented. Finally, the notation is include in appendix \ref{app:notation}.

\section{Gravitational domain wall solutions}

Gravitational domain walls are solutions to the Einstein-Hilbert equations coupled with a self-interacting scalar field,
\begin{equation}
G_{\hat{\mu}\hat{\nu}}+g_{\hat{\mu}\hat{\nu}}\Lambda = \pa{\hat{\mu}}\phi\pa{\hat{\nu}}\phi - g_{\hat{\mu}\hat{\nu}}\left(\frac{1}{2}\pa{\hat{\sigma}}\phi\pa{}^{\hat{\sigma}} \phi +V(\phi)\right).
\end{equation}
such that the potential of self-interaction has (at least) a couple of degenerated minima, and the scalar field interpolates between two consecutive minima. Since one is interested in solutions with a plane-parallel symmetry,  the scalar field depends only on the extra-dimension, $\xi$ \cite{Vilenkin-book}.

It is well known that the most general metric satisfying these conditions is,
\begin{equation}
g_{\hat{\mu}\hat{\nu}} = e^{2A(\xi)}\left[-dt_{\hat{\mu}}dt_{\hat{\nu}} + e^{2B(t,\xi)}dx^i_{\hat{\mu}}dx^i_{\hat{\nu}}\right]+e^{2C(\xi)}d\xi_{\hat{\mu}}d\xi_{\hat{\nu}}.\label{general-metric}
\end{equation}
Static and dynamical configurations can be found by considering $B$ as $B(\xi)$ or $B(t)$ respectively.
A bunch of gravitational domain wall solution might be found in the literature, and they can be generalized to various dimensions \cite{MPS}. Additionally, via scaling a thickness parameter can be introduced into the solutions, and families of solutions might be found \cite{GMP}. 

In the following the restriction to five dimensions is assumed, and only static solutions are considered.

\subsection{Kink  wall}

The kink domain wall is the simplest generalization of the solitonic $Z_2$-kink found in a 1+1-dimensional model \cite{Vilenkin-book,Goetz}. 

This solution can be achieved by setting in (\ref{general-metric}),
\begin{align}
A(\xi) = & -\frac{2}{3}\delta\left[\ln\left(\cosh\left(\frac{\alpha\xi}{\delta}\right)\right)+\frac{1}{4}\tanh^2\left(\frac{\alpha\xi}{\delta}\right)\right],\label{A-KDW}\\
  B(\xi)=&C(\xi)=0,
\end{align}
and  scalar field,
\begin{equation}
\phi(\xi) = \sqrt{3\delta}\tanh\left(\frac{\alpha\xi}{\delta}\right),\label{phi-KDW}
\end{equation}
where $\delta$ parametrizes the thickness of the wall, and $\alpha$ is related to the cosmological constant, which for this model is negative.
It will be shown in the next section that the self-interacting potential of the scalar field is not important for the purpose of fermion localization, and therefore its explicit form is omitted.

\subsection{Sine-Gordon wall}

The sine-Gordon domain wall is also the generalization of the 1+1-dimensional sine-Gordon model \cite{Vilenkin-book}. %%%%% add reference. 
 Since the self-interacting potential has an infinite number of degenerated vacua, there are different static solutions which are absent in a $Z_2$-model, such as the kink-kink solution.

In this case, the functions in (\ref{general-metric}) are
\begin{align}
A(\xi)=& -\delta\ln\left(\cosh\left(\frac{\alpha\xi}{\delta}\right)\right),\label{A-SGDW}\\
B(\xi)=&C(\xi)=0,
\end{align}
and for,
\begin{equation}
\phi(\xi)= \sqrt{3\delta}\arctan\left(\sinh\left(\frac{\alpha\xi}{\delta}\right)\right).\label{phi-SGDW}
\end{equation}
This is the single kink-like solution, where the scalar field interpolates between a couple of consecutive minima of the potential.

\subsection{Asymmetric wall}

Although  it has been shown that most  asymmetric domain walls do not localize gravity \cite{Castillo-Felisola:ULA,CastilloFelisola:2004eg}, it is known that they could emulate a gravitational attraction between particles on the wall. Thus, localization of fermions (and gauge fields lately) could result in a model where the fundamental content of the standard model lies on the brane, whilst four dimensional gravity is an effect of the embedding geometry of the brane \cite{Carter:2001af}.

Nevertheless, an example of asymmetric domain wall which does localize gravity has been found \cite{CastilloFelisola:2004eg}, with
\begin{equation}
A(\xi) = \alpha\xi-\delta\exp\left(-2 e^{-\frac{\beta\xi}{\delta}}\right) +\delta \Ei\left(-2 e^{-\frac{\beta\xi}{\delta}}\right),\label{A-ADW}
\end{equation}
\begin{equation*}
B(\xi)=C(\xi)=0,
\end{equation*}
and  scalar field,
\begin{equation}
\phi(\xi) = 2\sqrt{3\delta}\left(\exp\left(- e^{-\frac{\beta\xi}{\delta}}\right) - \varepsilon\right).\label{phi-ADW}
\end{equation}
Here the function $\Ei(\xi)$ is the exponential integral, defined by
\begin{equation*}
\Ei(x) = \int_{-\infty}^x dt \frac{e^t}{t},
\end{equation*}
and $\varepsilon$ is a shifting constant, which determines the asymptotic behavior of the fermionic modes.

\subsection{Double walls}

Double domain walls are a parametric family of solutions, whose energy density  peaks in two $\xi=$constant regions inside the wall thickness, and are a slight generalization of the thick brane of the Randall-Sundrum model presented in \cite{Gremm:1999pj}. These solutions are given by
\begin{equation}
  A(\xi)=C(\xi) = -\frac{1}{2s}\ln\left(1+(\alpha\xi)^{2s}\right),
\end{equation}
and
\begin{equation*}
  B(\xi)=0,
\end{equation*}
together with,
\begin{equation}
\phi(\xi) = \frac{\sqrt{6s-3}}{s}\arctan\(\alpha\xi\)^s.
\end{equation}
In order for $\phi$ to interpolate between to minima of the potential, $s$ must be odd. 

Fermion localization on this family of solutions has been studied  in a previous paper. Readers interested in a more exhaustive analysis of these domain walls are referred to \cite{CastilloFelisola:2010xh}.

\section{Fermion Localization on Static $D$-branes}\label{sec:loc}

In the following, thick-branes (or gravitational domain walls) are considered. The formal relation between  them might be established through a distributional limit \cite{GMP}.

Using the equations of motion coming from  the five dimensional  Dirac Lagrangian  on curved space-time with a Yukawa interaction,
\begin{equation}
  \left(-\not\!\nabla-M+\lambda\mathcal{P}(\phi)\right)\Psi=0,\label{Dirac-lag}
\end{equation}
and the decomposition of the five dimensional spinor in terms of four dimensional chiral ones, one gets the profile equations \cite{Kehagias:2000au,MPT}. These equations are Schr\"odinger-like, and their quantum mechanical potentials depend on the four dimensional chiral condition and Dirac equation. In the following, a brief review of the conditions that should be satisfied for (chiral) fermions to localize are shown, but a detailed analysis is presented in \cite{CastilloFelisola:2010xh}. We are only going to consider flat domain walls, i.e., $B(\xi=0)=0$.

\subsection{Massless 5D Fermions}

When  four dimensional chiral fermions, $\psi_\pm$, satisfy,  $$\not\! \partial^{(4)}\psi_\pm=0,$$
the profile equations are (see for example \cite{Kehagias:2000au,MPT}),
\begin{equation}
f'_\pm + A'f_\pm \mp \lambda\mathcal{P}(\phi)e^C f_\pm =0,
\end{equation}
where $f_\pm$ are the profiles for $\psi_\pm$. Their solutions are,
\begin{equation}
  f_\pm \propto e^{-A\pm \lambda\int d\xi\;\mathcal{P}(\phi)e^C}.\label{prof-flat-massless4d}
\end{equation}

However, four dimensional massive fermions mix chiralities, and the profile equations are coupled. Through a series of changes of variables, $\xi'=\int d\xi e^{-A+C}$ and $f_\pm \mapsto e^{-2A}u_\pm,$, the equations might be decoupled, resulting in a Sch\"odinger-like equation,

\begin{equation}
\left[-\partial^2_{\xi'} + V^\pm_{qm}\right]u_\pm = m^2 u_\pm,
\end{equation}
where $$f_\pm \mapsto e^{-2A}u_\pm,$$ and 
\begin{equation}
V^\pm_{qm} = \left(\lambda \mathcal{P}(\phi)e^A\right)^2 \pm \partial_{\xi'}\left(\lambda \mathcal{P}(\phi)e^A\right).\label{flat-Vqm}      
\end{equation}

\subsection{Massive 5D Fermions}

When a  five dimensional fundamental Dirac mass, $M$, is added, it can be shown that the above equations, change by substituting 
\begin{equation}
   \lambda\mathcal{P}(\phi) \to \lambda\mathcal{P}(\phi)-M.\label{poly-subs}
\end{equation}           

From (\ref{prof-flat-massless4d}), it is clear that after the substitution (\ref{poly-subs}), the localization of the positive profile, $f_+$, is favored in detriment of the negative one, $f_-$. 
This opposite effect serves as argument to suppose that regions (or a region) of the moduli space where both chiralities are localized exist(s).
In the case of a massive fermion, the effect enters through the quantum mechanical potential (\ref{flat-Vqm}), which is less intuitive to analyze.

Nonetheless, it is well known that adding a fundamental mass, $M$, in odd dimensional space-time gives a parity anomaly\cite{Niemi:1983rq,Redlich:1983dv}, which vanishes when a sign function on $\phi$ multiplies $M$\cite{Grossman:1999ra}, i.e. the substitution is,
\begin{equation}
   \lambda\mathcal{P}(\phi) \to \lambda\mathcal{P}(\phi)-M\sgn(\phi).\label{poly-subs2}
\end{equation}

\section{Analysis}

One might constraint the parameter space using analytic methods. In the following, the polynomial function, $ \mathcal{P}(\phi) $, will be restricted to a monomial, 
\begin{equation}
  \mathcal{P}(\phi) = \phi^n,
\end{equation}
to avoid the introduction of more free parameters.

In order to assure that one of the chiralities is localized, a constraint is found for the coupling constant, say,
\begin{equation}
f_\pm(\xi) \propto \exp\left({-A\pm\int d\xi\lambda\mathcal{P}(\phi)e^C}\right),\label{prof-constraint}
\end{equation}
when $M=0$. Additionally, in the simplest model a small coupling constant value does not allow localization of the positive chirality, independently of the domain wall considered. Thus, if both chiralities are expected to be localized, a new constraint on $\lambda$ is found.

For the numerical analysis of the moduli space, a Numerov algorithm together with a bisection method is used for finding the eigenvalue of the ground state of the Schr\"odinger-like equation for the profiles. The parameters are restricted to $\lambda\in [1.00,5.00]$ sliced in steps of $0.05$, and $M\in[0.00,1.00] $ in steps of $0.02$. These algorithms were implemented using SAGE \cite{sage}.

The existence of tachyonic modes has been used to rule out several models, since they are a signal of instabilities in the model, due to causality. Nonetheless the recent findings of OPERA\cite{:2011zb} could be interpreted as possible evidence of tachyonic behavior of neutrinos through matter. In the figures, the vertical axis takes one of the two possible values, 0 if the choice of parameters implies existence of tachyonic instabilities, while 1 means that the choice of parameters is phenomenologically plausible, {\it i.e.} tachyon-free.

\subsection{Kink domain wall}

Using the warp factor and the scalar field corresponding to the kink domain wall, i.e. (\ref{A-KDW}) and (\ref{phi-KDW}), in  (\ref{prof-constraint}),  one gets a lower bound constraint for the coupling constant, due to the negative chirality,
\begin{equation}
 \lambda > \frac{2\alpha}{3}\frac{1}{(3\delta)^{n/2}}.
\end{equation} 

This is the generalization of the constraints found in \cite{MPT}. A similar constraint was found in \cite{CastilloFelisola:2010xh} for the double wall scenario.

%%%%%%%%%% Sign function
\begin{figure}
\includegraphics[scale=.24]{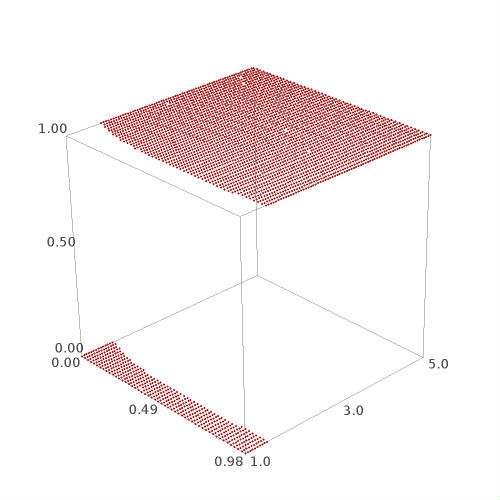}
\includegraphics[scale=.24]{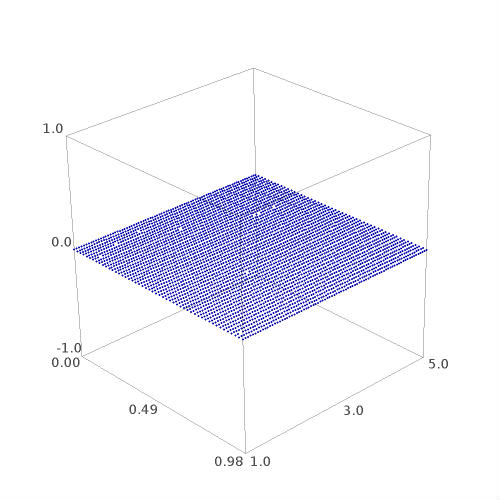}
\caption{Moduli space for kink domain walls for linear Yukawa coupling and sign function, for left- and right-handed fermions.}\label{fig:Kink-n_1}
\end{figure}

%% \begin{figure}
%% \includegraphics[scale=.24]{K-Vn-Mod++n_3.jpeg}
%% \includegraphics[scale=.24]{K-Vp-Mod+n_3.jpeg}
%% \caption{Moduli space for kink domain walls for cubic Yukawa coupling and sign function, for left- and  right-handed fermions.}\label{fig:Kink-n_3}
%% \end{figure}

%% \begin{figure}
%% \includegraphics[scale=.24]{K-Vn-Mod++n_5.jpeg}
%% \includegraphics[scale=.24]{K-Vp-Mod+n_5.jpeg}
%% \caption{Moduli space for kink domain walls for fifth Yukawa coupling and sign function, for left- and right-handed fermions.}\label{fig:Kink-n_5}
%% \end{figure}

In figure \ref{fig:Kink-n_1} it can be seen that the left-handed profile does localize itself, without the existence of tachyonic modes, for any value of $M$ if $\lambda>\lambda_c \sim 1.44$. On the other hand, the right-handed profiles are not localized in a tachyon free theory.

Unlike the $n=1$ case, for cubic and fifth scalar-fermion coupling the left chirality does not present tachyonic modes while right profiles inevitable present tachyons, for any value of $(\lambda,M)$.

\subsection{Sine-Gordon domain wall}

Using (\ref{A-SGDW}) and (\ref{phi-SGDW}) in (\ref{prof-constraint}), yields the constraint for the profile. However, integrating the argument of the exponential becomes highly non-trivial for monomials other than linear. Therefore, an approximation might be done,
\begin{equation}
 f_\pm(\xi)\propto \exp\left(\int d\xi \Upsilon(\xi)\right).\label{prof-approx}
\end{equation} 
Equation (\ref{prof-approx}) defines the function $\Upsilon$. Thus, the asymptotic behavior of $\Upsilon$ determines whether or not the exponential converges. Convergence demands, 
\begin{equation}
 \lambda > \alpha\left(\frac{2}{\sqrt{3\delta}\pi}\right)^n.
\end{equation}

The numerical analysis for this kind of domain wall assures that there is not a single point in the considered region of the moduli space which provides a tachyon free model of fermions.

\subsection{Asymmetric domain wall}

It was shown in \cite{MPT} that for linear fermion-scalar coupling, (\ref{prof-constraint}) can be integrated explicitly. However, for more general couplings this is not the case, and (\ref{prof-approx}) should be used.

Due to the asymmetry, the limit values of $A(\xi)$ and $\phi(\xi)$ as $\xi\to\pm\infty$ differ, and conditions for localization vary for each limit and chirality. From (\ref{prof-approx}) one get,
\begin{align}
 -\alpha+\beta\pm\lambda\left(2\sqrt{3\delta}(1-\varepsilon)\right)^n < & 0,\\
-\alpha\pm\lambda\left(2\sqrt{3\delta}\varepsilon\right)^n < & 0,
\end{align}
for $\xi\to+\infty$ and $\xi\to-\infty$ respectively, which in terms of the cosmological constants and for  negative chirality are,
\begin{align}
 \lambda > & \frac{\sqrt{\Lambda^+}}{\sqrt{6}(2\sqrt{3\delta}(1-\varepsilon))^n},\\
\lambda < & \frac{\sqrt{\Lambda^-}}{\sqrt{6}(2\sqrt{3\delta}\varepsilon)^n}.
\end{align}
If one demands the saturation of both equations, a constraint on $\varepsilon$ is found,
\begin{equation}
 \frac{1}{\varepsilon} = 1+\left(\frac{\Lambda^+}{\Lambda^-}\right)^{\frac{1}{2n}},
\end{equation} 
and the critical value of the coupling turns out to be,
\begin{equation}
 \lambda_* = \frac{1}{\sqrt{6}(2\sqrt{3\delta})^n}\left((\Lambda^+)^{\frac{1}{2n}}+(\Lambda^-)^{\frac{1}{2n}}\right)^n.
\end{equation} 

\begin{figure}
\includegraphics[scale=.24]{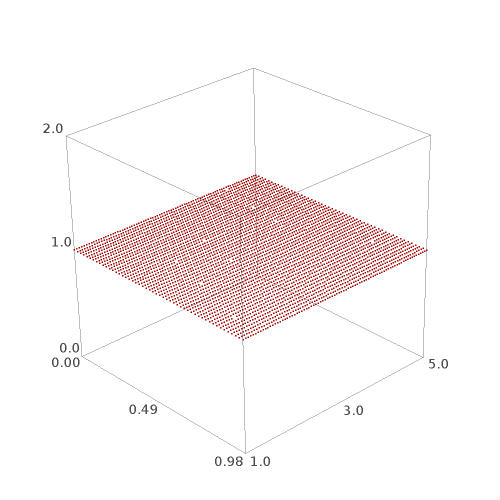}
\includegraphics[scale=.24]{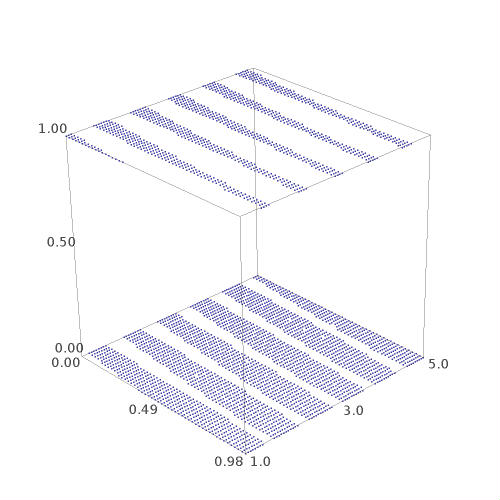}
\caption{Moduli space for asymmetric domain walls for linear Yukawa coupling, for left- and right-handed fermions.}\label{fig:Asym-n_1}
\end{figure}

\begin{figure}
\includegraphics[scale=.24]{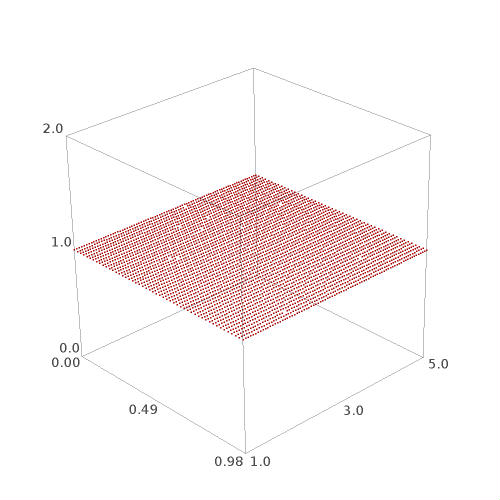}
\includegraphics[scale=.24]{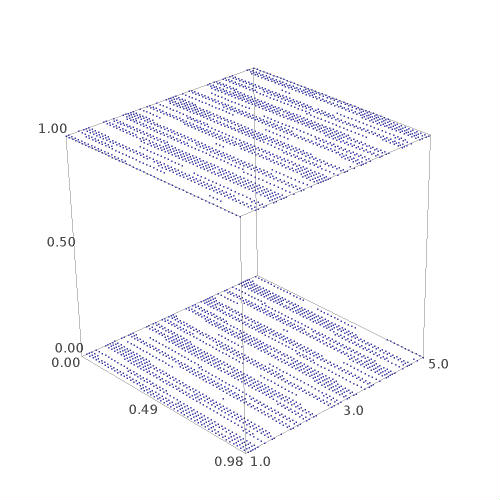}
\caption{Moduli space for asymmetric domain walls for cubic Yukawa coupling, for left- and right-handed fermions.}\label{fig:Asym-n_3}
\end{figure}

\begin{figure}
\includegraphics[scale=.24]{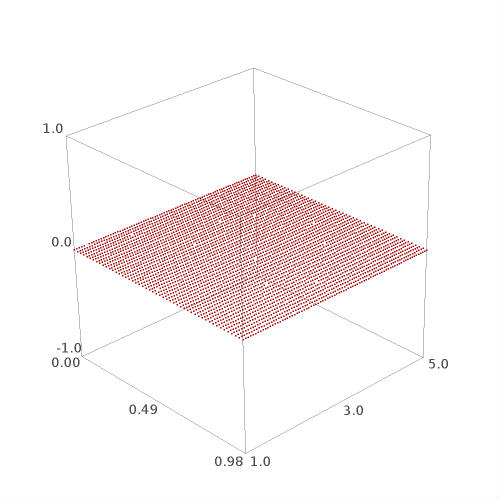}
\includegraphics[scale=.24]{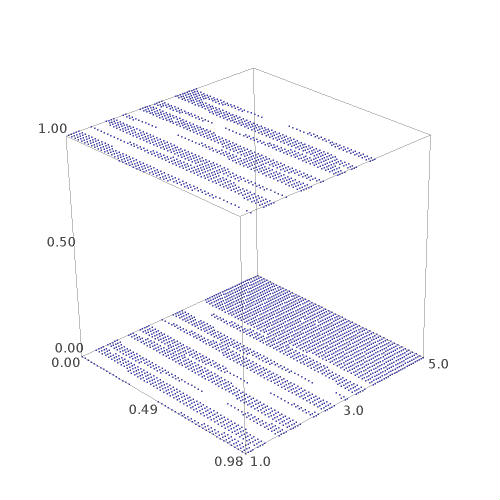}
\caption{Moduli space for asymmetric domain walls for fifth Yukawa coupling, for left- and right-handed fermions.}\label{fig:Asym-n_5}
\end{figure}

The numerical analysis of the moduli space for this asymmetric domain walls gives a very interesting feature. As can be seen in figures \ref{fig:Asym-n_1}, \ref{fig:Asym-n_3} and \ref{fig:Asym-n_5}, the left handed mode is either localized (for linear and cubic coupling) or unlocalized (fifth coupling). However, the right handed fermion presents a sort of ``band'' structure \footnote{Similar to the electronic bands which appears in condensed matter.}, depending on whether the model is tachyon-free (upper edge of the cube) or contains fermionic tachyonic modes.
%%%%%%%%%% No sign function

So far, the moduli space of these models has been separated into pieces where tachyons can or cannot exist.

%%%%%%%%%% Sign function
\begin{figure}
\includegraphics[scale=.24]{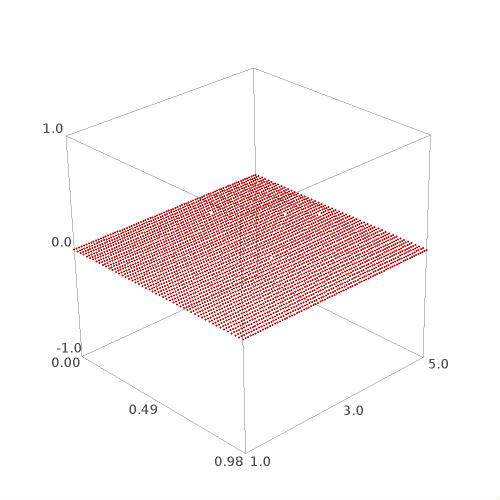}
\includegraphics[scale=.24]{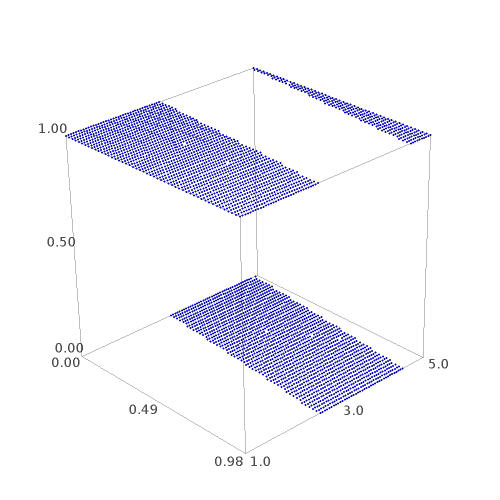}
\caption{Moduli space for asymmetric domain walls for linear Yukawa coupling and sign function, for left- and right-handed fermions.}\label{fig:Asym+n_1}
\end{figure}

%% \begin{figure}
%% %\includegraphics[scale=.24]{Asym-Vn-Mod+n_3a.jpeg}
%% \includegraphics[scale=.34]{Asym-Vp-Mod+n_3.jpeg}
%% \caption{Moduli space for asymmetric domain walls for cubic Yukawa coupling and sign function, for  right-handed fermions.}\label{fig:Asym-n_3}
%% \end{figure}

%% \begin{figure}
%% %\includegraphics[scale=.24]{Asym-Vn-Mod+n_5a.jpeg}
%% \includegraphics[scale=.34]{Asym-Vp-Mod+n_5.jpeg}
%% \caption{Moduli space for asymmetric domain walls for fifth Yukawa coupling and sign function, for right-handed fermions.}\label{fig:Asym-n_5}
%% \end{figure}

When the sign function, which avoids a parity anomaly, is turned on, the left handed fermion presents always a tachyonic mode. On the other hand, the localization of the right handed profile for $n=1$ has a band structure, where tachyonic modes can or cannot be avoided in the phenomenological model. For any other value of $n$ no tachyon free model can be constructed. 

\section{Discussion}

Nowadays, due to the results of the OPERA experiment, physicists should ask themselves whether or not to  model the universe allowing tachyonic modes. This new possibility opens the Pandora's box of causality, and of course new theories are been developed in this direction.

Here, by considering  brane-world scenarios with an extra dimension, the simplest moduli space of parameters has been studied, and the numerical analysis  has  shown that the moduli space of parameters can be split into areas whose physical models are tachyon free and areas where this feature cannot be achieved.

Several kinds of domain wall models were considered, and it seems that the most interesting structure of the moduli space is the one of the asymmetric domain wall, which presents what looks like a band structure, similar to the electronic structure of condensed matter. Although higher fermion-scalar couplings were studied, the numerical results show that the most plausible scenario is the linear coupling among these fields.

The sine-Gordon domain wall presents a moduli space which does not permit (in the considered region) to model physics without tachyonic fermions. Unless more data confirm the existence of the tachyonic modes in the neutrino sector, the sine-Gordon domain wall scenarios must be ruled out for physical models.

In kink domain walls, the $n=1,3,5$ right handed fermions require the existence of tachyonic  modes, while the left handed one is tachyon free, except for the case $n=1$, which additionally requires $\lambda>\lambda_c\sim 1.44$. Therefore, for $\lambda<\lambda_c$, both fermion chiralities present tachyon modes.

%Although the numerical analysis constrains strongly the moduli space of parameters, for tachyon free physical models, some phenomenological point might be dressed out.

Some phenomenology can be read from the numerical analysis, despite the fact that the constraints on tachyon free models are strong. However, when models are constructed it is important to have lines which allow to rule them out.

Depending on the algebraic form of the quantum mechanical potentials, $V_{qm}^\pm$, the localization of the left handed chirality is favored in detriment of the right  handed one,  due to the appearance of a barrier near the position of   the $D$3-brane.  Since light right handed fermions tend to propagate out of the brane, one might suggest that  the OPERA results might be interpreted as propagation of right handed neutrinos through the warped extra dimension, as suggest in \cite{Hollenberg:2009qf} as an explanation of the LSND neutrino anomaly.

\section*{Acknowledgement}

We would like to thank V. Shama and J. Zamora for helpful discussions and comments. O. C-F. thanks Swansea University for there hospitality during part of the development of this work. This work was supported in part by MECESUP2 (Chile) under Grant No. FSM0605-D3008, and Fondecyt Project No. 11000287.

% LocalWords:  chiralities chirality Yukawa qm  brane tachyons

\appendix

\section{Conventions and Notations}\label{app:notation}

Through the manuscript, the metrics have signature mostly positive. Since the tetrads formalism is used extensively, distinction between flat and curved coordinates by Latin and Greek indices. Moreover, hated indices run over the whole spacetime whilst unhated ones run over a hypersurface restriction, i.e., on the coordinates parallel to the topological defect.

The gamma matrices are defined in the tangent space, and they satisfy the Clifford algebra,
\begin{equation}
  \left\{\GAM{a},\GAM{b}\right\}=2\eta^{\hat{a}\hat{b}}\mathds{1}.\label{Cliff-alg}
\end{equation}
In even dimensions one can define the chirality matrix $\ga^*$, satisfying the properties
\begin{itemize}
\item $\left\{\GAM{a},\ga^* \right\}=0$.
\item $(\ga^*)^2=\mathds{1}$.
\end{itemize}
From this, the projector operators,
\begin{equation}
  P_\pm= \frac{\mathds{1}-\ga^*}{2},
\end{equation}
are both non-trivial.

On the other hand, odd dimensional Clifford algebras are constructed by using the gamma matrices of the co-dimension one spacetime, via $\GAM{a} = \(\gam{b},\ga^*\)$. These odd dimensional Clifford algebras have a trivial projectors $P_\pm$, and therefore chiral fermions cannot be defined.

In any dimension one may define a set of generators of the Lorentz algebra, constructed with the elements of the Clifford algebra (\ref{Cliff-alg}). These generators of the Lorentz algebra are,
\begin{equation}
  \J^{\hat{a}\hat{b}}=-\frac{\imath}{4}\left[\GAM{a},\GAM{b}\right].
\end{equation}

When the Dirac-Feynman slash notation is used it must be interpreted as,
\begin{equation}
  \fy\partial=\VIN{\mu}{a}\GAM{a}\PA{\mu}.
\end{equation}

The covariant derivative for gauge theories is defined by,
\begin{equation}
  \NAB{\mu}=\PA{\mu}-\imath g A_{\hat{\mu}}-\imath \Omega_{\hat{\mu}},
\end{equation}
with $\Omega$ the gravitational connection, which is related to the Christoffel connection for integer spin fields, and with the spin connection for semi-integer spin fields. Clearly the Dirac-Feynman slash notation can be used with the covariant derivative,
\begin{equation}
  \fy\nabla=\VIN{\mu}{a}\GAM{a}\NAB{\mu}=\VIN{\mu}{a}\GAM{a}\left(\PA{\mu}-\imath g A_{\hat{\mu}}-\imath \Omega_{\hat{\mu}} \right).
\end{equation}

\begin{comment}
  If the vielbein have a dependency on the 4-dimensional coordinates the distinction between the 4-dimensional restriction and the component of the total 5-dimensional one is made by denoting with a hat the 5-dimensional one, i.e.,  if the metric with line element 
  \begin{small}
    \begin{equation}
      ds^2(g)=e^{2A(\xi)}\left[- dt^2 + e^{2B(r)}dr^2+r^2d\Omega^2\right] + d\xi^2,
    \end{equation}
  \end{small}
  the 5-dimensional vielbein are,
  \begin{align}
    \hat{e}^0=e^{A}\df t,\quad &\hat{e}^1= e^{A+B}\df r,\quad \hat{e}^2=r e^A\df\theta,\nonumber\\ 
    \hat{e}^3= r\sin\theta e^A\df\varphi \quad & \text{and}\quad \hat{e}^5={e}^5=\df\xi,
  \end{align}
  whilst the 4-dimensional restriction vielbein are,
  \begin{align}
    {e}^0=\df t,\quad & {e}^1= e^{B}\df r,\nonumber\\
    {e}^2=r\df\theta \quad\text{and}&\quad {e}^3= r\sin\theta\df\varphi,
  \end{align}
\end{comment}

\bibliographystyle{unsrt}
\bibliography{References.bib}

\end{document}